# CHOICE SETS AND SMART CARD DATA IN PUBLIC TRANSPORT ROUTE CHOICE MODELS: GENERATED VS. EMPIRICAL SETS

Georges Sfeir
*University of Leeds, Leeds, United Kingdom, g.sfeir@leeds.ac.uk*
Filipe Rodrigues
*Technical University of Denmark, Kongens Lyngby, Denmark, rodr@dtu.dk*
Ravi Seshadri
*Technical University of Denmark, Kongens Lyngby, Denmark, ravse@dtu.dk*
Carlos Lima Azevedo
*Technical University of Denmark, Kongens Lyngby, Denmark, climaz@dtu.dk*

**Abstract**: This study evaluates path sets generation for route choice models in multimodal public transportation networks, using both conventional (network algorithms) and empirical (smart card data driven) methods. While the empirical approach can present limitations with a short observation period, it improves substantially with more data, offering a computational efficiency advantage over conventional methods. In such approach, while incorporating real-world delays increased travel time variability, it still aligned with planned travel times, and relaxing access/egress assumptions further enhanced coverage. Work is undergoing on the evaluation of the impact of different choice sets in bias and efficiency of route choice parameter estimates.

______________________________________________________________________

## 1. INTRODUCTION AND BRIEF BACKGROUND

Public transport route choice models are crucial for understanding travel behavior and network design. In their classical form, they consist of two main components: Choice Set Generation (CSG, identifying all possible route alternatives) and choice modeling. CSG accuracy can significantly affect model estimation and demand forecasting (Swait & Ben-akiva, 1987). Classic methods such as k-shortest path and link elimination, to name a few, rely on topological criteria rather than passenger preferences. Consequently, some generated routes may not reflect actual choices (Gentile & Noekel, 2016). Effective CSG should balance coverage and precision for better parameter estimate (Zimmermann & Frejinger, 2020). However, defining relevant routes is complex and subjective, which has recently motivated alternative approaches to tackle (Duncan et al., 2023) or avoid CSG (Fosgerau et al., 2022). Recent studies have leveraged Smart Card (SC) data from Automated Fare Collection systems to generate empirical choice sets (Arriagada et al., 2022), capturing nearly all travelers and producing more realistic choice sets. However, this approach excludes unchosen but considered routes and often relies on selected observation periods, which may not capture all relevant paths. A longer observation period can generate a more exhaustive set but the required size, efficiency and accuracy gains remains uncertain. This paper studies CSG for a large multimodal public transport network using both conventional methods and SC data. It evaluates their computational performance, coverage, and composition.

## 2. CHOICE SETS GENERATION

We focus on the multimodal public transport network in the East Great Belt area of Denmark including the Greater Copenhagen Region. Under the Danish SC system (Rejsekort) passengers must tap-in at transfer and tap-out. The Rejsekort system covers all public transport modes (buses, trains, and metros), and travel zones in Denmark. Each Rejsekort transaction stores information on the type of transaction, time and location of the transaction, type of the card, and anonymized card ID.

### 2.1. Conventional choice set

GTFS data and road network from OpenStreetMap (OSM), were used to construct a public transport graph and compute path attributes: in-vehicle travel time, waiting time, transfers, walking time, and path size. The graph consists of vertices representing bus stops, train/metro stations, and road network nodes, connected by four edge types (bus, train, metro, walk). The case study includes 136,031 vertices and 3,513,457 edges, comprising 11,419 bus stops, 1,425 bus lines, 244 train/metro stations, 46 train/metro lines, 124,368 access nodes, and 784,603 road segments. In-vehicle travel time was averaged across all service lines on a segment. Walking time was computed with a constant speed of 4 km/h. Waiting time was determined based on service line frequencies per segment. Waiting times were based on service line frequencies on each route segment.

With the multimodal network constructed, a one-day smart card dataset (19/09/17) was processed, extracting all observed stop-to-stop trips, resulting in 88,700 unique OD pairs after data cleaning. A conventional choice set was then generated using four CSG algorithms: k-shortest path, link elimination, labelling, and simulation, yielding paths for 86,128 OD pairs. The final choice set contained 39.47% of OD pairs with a single path and an average of 2.60 alternatives per OD and resulted on a high coverage of 98.88%. The CSG computation required 2 weeks on a 2.6GHz CPU and 196GB RAM linux machine.

### 2.2. Observed/Empirical choice set

The same SC data was selected, and observed paths were generated on the multimodal network for each OD pair recorded. Travel time for each path was computed as the average observed travel time, incorporating delays. To be consistent with the conventional choice set based on planned travel times, observed paths were matched with GTFS data to extract planned travel times (instead of realized ones), determine the number of transfers (including hidden transfers), and compute path-size factors. The empirical choice set showed that 66.36% of OD pairs had only one alternative, with an average of 1.61 alternatives per OD. Expanding the dataset from 1 to 20 weekdays reduced single-alternative OD pairs to ~46% and increased the average alternatives per OD from 1.61 to 6.35 (Figure 1 and 2). No plateau effect was observed, suggesting additional data may further refine the CSG process. In terms of computational performance, it took less than 5 minutes to generate the choice set with observed travel times for 20 days of data, using the same machine as in the conventional method. When matching the observed paths with GTFS data to obtain planned travel times, the process took approximately 2 days. However, this is still significantly more efficient than the conventional method.

An effective CSG approach should avoid generating paths with an excessive number of transfers compared to the path with the minimum transfers for each OD pair, as these paths are likely irrelevant. The additional number of transfers per OD pair serves as an indicator of the choice set quality (Rui, 2016). The ECDF in Figure 2-a shows that empirical choice sets generated with 10 days or fewer contain fewer irrelevant paths (i.e., paths with excessive transfers) than the conventional choice set. However, as the number of days increases to 15 and 20, the proportion of paths with excessive transfers in the empirical choice set becomes comparable to that of the conventional choice set. Figure 2-b presents the ECDF of the standard deviation of total travel time, comparing planned travel times in the conventional choice set with observed travel times in the empirical choice sets. The results indicate that both approaches generate paths with relatively low variability in travel time. However, empirical choice sets exhibit slightly higher variability due to real-world delays, disruptions, and operational inconsistencies captured in smart card data, whereas the conventional choice set is based solely on scheduled times. Additionally, increasing the number of days considered in the empirical choice set leads to a gradual rise in travel time variability, reflecting a wider range of network conditions such as congestion and service disruptions. To ensure a direct comparison of planned travel time variability, Figure 2-c presents the ECDF of planned total travel times for both choice sets, excluding the impact of real-world delays. The results show that both approaches exhibit low travel time variability across OD pairs, with only a slight increase in variability as the number of days in the empirical choice set increases. This suggests that including more days in the empirical choice set expands the set with additional observed and feasible alternative paths with different planned travel times. However, despite this increase, the planned travel time variability in the empirical choice sets remains similar to that of the conventional choice set. This finding indicates that although empirical choice sets are based on observed passenger behavior, their planned travel time characteristics do not significantly diverge from those of the conventional choice set, even as more days are considered.

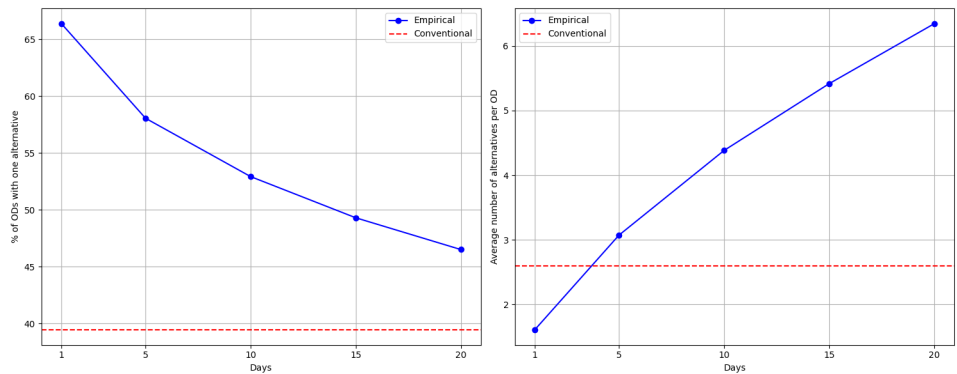

**Figure 1. Empirical vs. Conventional Methods: Number of Alternatives Across Days**

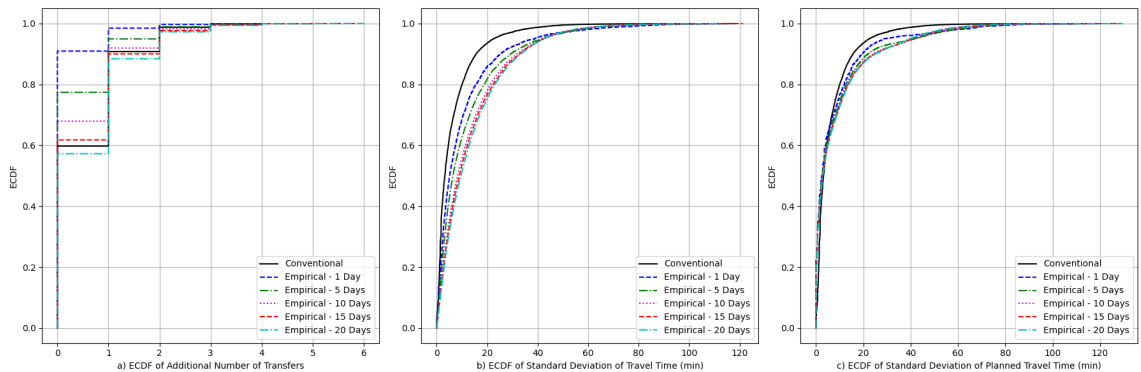

**Figure 2. Empirical vs. Conventional Methods: Transfers and Travel Times Across Days**

## 2.3. Sensitivity to access and egress alternatives

To better reflect passenger behavior, the empirical choice set was expanded to include trips within a 500-meter radius of original boarding and alighting stops. This adjustment significantly improved choice set quality. With one day of data, OD pairs with only one alternative dropped from 66% (stop-to-stop) to 40% with the 500-meter radius. After 20 days, this further reduced to 5% (vs. 46% in the stop-to-stop case). The average number of alternatives per OD also increased, reaching 11.0 after 20 days compared to 6.35 in the original method. Figure 3 indicates that while observed travel times in empirical choice sets show slightly higher variability due to real-world disruptions, planned travel times remain stable. The 500-meter approach also maintains fewer excessive-transfer paths than the conventional choice set, ensuring better behavioral realism without distorting travel conditions.

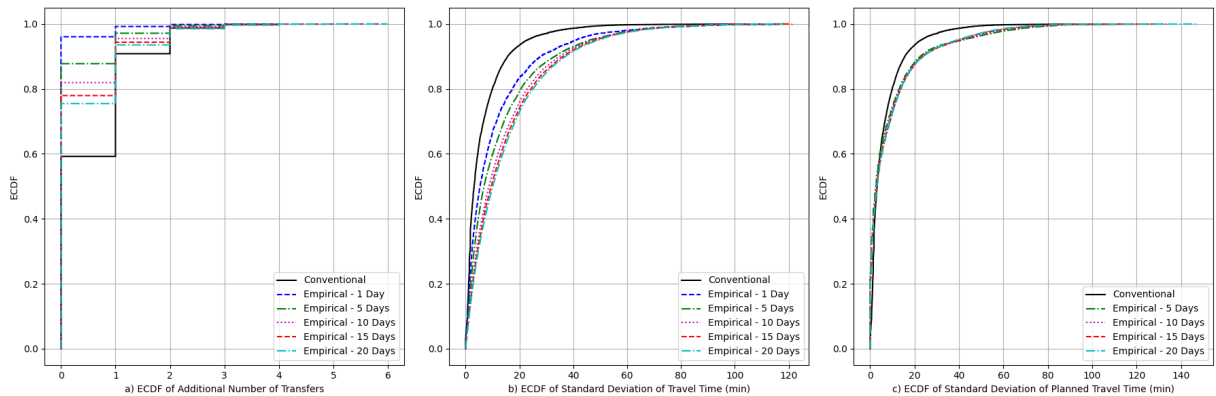

**Figure 3. Transfers and Travel Times Across Days – 500-meter radius**

## 3. DISCUSSION

Our study explored the generation of public transport choice sets using both conventional and empirical methods. The empirical approach showed several feature limitations in a low period of observation setting, yet minimized as more days of data were included. However, this approach offered a significant computational efficiency advantage. While including real-world delays and disruptions in the empirical choice set increased travel time variability, the variability in planned travel times remained relatively consistent between both methods. The empirical set's ability to capture operational inconsistencies, such as congestion or service disruptions, may offer a more realistic reflection of passenger behavior without significantly diverging from scheduled travel times. We show that the relaxation of access and egress assumptions in using stop-to-stop data further enhanced the coverage and diversity of alternatives. Our analysis still stands on evaluation of the characteristics of choice set, and future research should explore its impacts on the bias and efficiency of route choice parameter estimates.

## REFERENCES

Arriagada, J., Munizaga, M. A., Guevara, C. A., & Prato, C. (2022). Unveiling route choice strategy heterogeneity from smart card data in a large-scale public transport network. *Transportation Research Part C: Emerging Technologies*, *134*. https://doi.org/10.1016/j.trc.2021.103467

Duncan, L. C., Watling, D. P., Connors, R. D., Rasmussen, T. K., & Nielsen, O. A. (2023). Choice set robustness and internal consistency in correlation-based logit stochastic user equilibrium models. *Transportmetrica A: Transport Science*, *19*(3). https://doi.org/10.1080/23249935.2022.2063969

Fosgerau, M., Paulsen, M., & Rasmussen, T. K. (2022). A perturbed utility route choice model. *Transportation Research Part C: Emerging Technologies*, *136*. https://doi.org/10.1016/j.trc.2021.103514

Gentile, G., & Noekel, K. (2016). *Modelling Public Transport Passenger Flows in The Era of Intelligent: COST Action TU1004 (TransITS) Transport Systems* (10th ed., Vol. 1). Cham: Springer International. Print. Springer Tracts on Transportation and Traffic.

Rui, T. (2016). *Modeling route choice behaviour in public transport network*. PhD. Thesis, MIT.

Swait, J., & Ben-akiva, M. (1987). Incorporating random constraints in discrete models of choice set generation. *Transportation Research Part B: Methodological*, *21*(2), 91–102.

Zimmermann, M., & Frejinger, E. (2020). A tutorial on recursive models for analyzing and predicting path choice behavior. *EURO Journal on Transportation and Logistics*, *9*(2). https://doi.org/10.1016/j.ejtl.2020.100004

## ACKNOWLEDGMENT

The research leading to these results has received funding from: 1) the Horizon Europe Framework Programme under the Marie Skłodowska-Curie Postdoctoral Fellowship MSCA2021-PF-01 project No 101063801; and 2) the SORTEDMOBILITY project which is supported by the European Commission and funded under the Horizon 2020 ERA-NET Cofund scheme under grant agreement N. 875022